\documentclass[12pt]{article}

\usepackage{amsmath}
\usepackage{graphicx}

\title{Bottomonium  $\Upsilon(5S)$
 decays into $BB$ and $BB\pi$}

\author{  Yu.A.Simonov, A.I.Veselov\\State Research
Center\\Institute of Theoretical and Experimental Physics, \\
Moscow, 117218 Russia}

 \date{}
\newcommand{\beq}{\begin{eqnarray}}
 \newcommand{\eeq}{\end{eqnarray}}
\newcommand{\be}{\begin{equation}}
 \newcommand{\ee}{\end{equation}}
\author{  Yu.A.Simonov, A.I.Veselov\\State Research
Center\\Institute of Theoretical and Experimental Physics, \\
Moscow, 117218 Russia}
 \date{}

\def\fun#1#2{\lower3.6pt\vbox{\baselineskip0pt\lineskip.9pt
\ialign{$\mathsurround=0pt#1\hfil ##\hfil$\crcr#2\crcr\sim\crcr}}}

\newcommand{{\SD}}{\rm SD}
\newcommand{{\Lc}}{\mathcal{L}}
\newcommand{{\Mc}}{\mathcal{M}}

\newcommand{\vep}{\mbox{\boldmath${\rm p}$}}
\newcommand{\veq}{\mbox{\boldmath${\rm q}$}}

\newcommand{\vek}{\mbox{\boldmath${\rm k}$}}

\begin{document}
\maketitle
\begin{abstract}
Two- and three-body decays of $\Upsilon (5S)$ into $BB,
$$ BB^*, $ $ B^*B^*,$ $  B_sB_s, $ $B_sB_s^*, $ $B_s^* B_s^*$ and
$BB^*\pi, B^*B^*\pi$ are evaluated using the theory, developed earlier  for
dipion bottomonium transitions. The theory contains only two parameters, vertex
masses $M_{br}$ and $M_\omega$, known from dipion spectra and width. Predicted
values of $\Gamma_{tot}(5S)$ and six partial widths $\Gamma_k(5S), k=BB,
BB^*,...$ are in agreement with experiment. The decay widths $\Gamma_{5S}(\pi
BB^*)$ and $\Gamma_{5S}(\pi B^*B^*)$ are also calculated and found to be of the
order of 10 keV.


\end{abstract}

\section{Introduction}
The experimental data \cite{1} on dipionic and dikaonic
transitions of $\Upsilon(5S)$, together with the earlier data on
$BB$ and total width \cite{2} present an interesting challenge to
 theorists, asking for explanation of several unexpected
features:  1) dipionic widths of $\Upsilon(5S)$ are almost $10^3$
larger than of  $\Upsilon(nS), n\leq 4;$ 2) the  $BB, BB^*$ etc.
partial widths of $\Upsilon(5S)$ are large, and $\Gamma_{tot} (5S)
\approx 110$ MeV \cite{2}; 3) a hierarchy exists
 in $\Gamma_k (5S), k=1,2,..6$, e.g. $\Gamma_{BB} < \Gamma_{BB^*}
 <\Gamma_{B^*B^*}.$

 In addition there is an open problem of finding the decay width
 into the $BB^*\pi, B^* B^*\pi$  channels,  not yet known
 experimentally.

 From theoretical point of view a calculation and comparison of
 two types of decays (6 channels for $BB$ and 2 channels for
 $BB\pi$ ) is important, because it allows to estimate two
 independent types of light $q\bar q$ creation vertices:
 $L=M_i\int\bar \psi(x) \psi(x) d^4x,~~ i=\omega, br$, where $M_\omega$ governs pair creation without pions and was
 found previously \cite{3,4,5,6} (both experimentally and
 theoretically) to be of the order of 1 GeV and $M_{br}$, which is
 responsible for pair creation with additional pion emission,
 $M_{br}$  being of the order of $f_\pi=0.093 $ GeV \cite{3,4,5,6}. It is clear, that $M_\omega$
  enters in the  6 channels of the $BB$ type  and
 $M_{br}$ enters in two $BB\pi$ channels, therefore the resulting
 widths would strongly differ from each other.  It is purpose of
 the present paper to calculate all 8 widths and compare results
 with existing experimental data, answering in this way to the
 points 2) and 3) stated above.

 In doing so we shall  use the values of $M_\omega$ and $M_{br}$ from the analysis of $\Upsilon (nS)$ decays with $n\leq 4$
  and hence no new parameters will be
 involved.

 The paper is organized as follows. In the  next section the
 general equations from \cite{3,4,5,6} are written for the
 widths $\Gamma_k(5S)$  and
 evaluated, using realistic one-channel wave-functions of
 $\Upsilon (nS)$,  calculated earlier. In section 3 the $\pi BB$
 widths are written and calculated, and in the concluding section
 results are discussed and perspectives are outlined.

 \section{The $BB$ decay channels of $\Upsilon (5S)$}

We numerate 6 channels of $BB$ type as follows; $k=1,2,..6$ for
$B\bar B,$ $ B\bar B^*, $ $B^* \bar B^*, $ $B_s\bar B_s,
$$B_s\bar B^*_s, $$ B_s^* \bar B_s^*$ respectively, and define the
partial widths $\Gamma_k$, which according to general formulas
(see e.g. Eq. (63) in \cite{3}, and Eq (28 ) in \cite{5}) \be
\Gamma_k(\Upsilon (5S)) \equiv \Gamma_k= \left(
\frac{M_\omega}{2\omega}\right)^2 \frac{p^3_kM_k}{6\pi
N_c}(Z_k)^2|J_{5_S} (p_k) |^2.\label{1}\ee

Here $\omega$ is the light quark average energy in $B$ or $B_s$ mesons,
$\omega_B=0.587$ GeV, $\omega_{B_s}=0.639$ GeV (see Appendix 1 of \cite{3}),
$Z_k$ are spin-isospin multiplicity factors, \be Z^2_1 = 2 Z^2_4 =1;~~ Z^2_2 =
2 Z^2_5 =4: ~~ Z^2_3= 2Z^2_6 =7\label{2}\ee $M_k$ is the doubled   reduced mass
in channel $k$, and $p_k$- the corresponding momentum. Finally, $J_{5S}(p)$ is
the overlap matrix element of  the wave functions $\Psi_5 $ of $\Upsilon (5S)$
and $\psi_B$ or $B$ mesons, \be \frac{p_i}{\omega} J_{5S} (p) =\int \frac{d^3
q}{(2\pi)^3} \bar y_{123} \Psi^*_5 (\vep +\veq) \Psi^2_B(\veq)\label{3}\ee
where $\bar y_{123}$ is the  $P$-wave vertex factor given in \cite{3,4,5,6},
$\bar y_{123} =q_i-\bar c p_i$,  $\bar c \ll 1$. In \cite{6} we have used for
$\Psi_5 (q)$ the realistic wave function for $\Upsilon (5S)$ obtained in
\cite{7} in the one-channel approximation, i.e. without influence of decay
channels, and expanded  in a series of oscillator functions for convenience. As
a result $\Psi_5(q)$ contains   4 real zeros as a function of $q^2$, which
yields $J_{5S} (p)$ strongly varying in region of physical values of
$p_k=0.48\div 1.26$ GeV. This makes $\Gamma_k$ to depend on the shape of
$\Psi_5 (q)$ and details of its approximation. At the same time one realizes,
that the same decay interaction due to $M_\omega\bar \psi\psi$ strongly mixes
the wave functions of  $n^3S_1$ and $n^3D_1$ with different $n$, so that the
$5S$ state gets (complex due  to open channels) admixture of $n'$ states with
$n'\neq 5$. The preliminary analysis shows that this contribution fills the
minima of wave function and overlap integral, yielding a smooth function
$J_{5S} (p)$. Therefore in the Table 1 we display the "averaged" values
$J^{(0)}_{5S} (p) =\exp \left(-\frac{p^2}{\Delta_5}\right) \bar I_{5S}$, with
$I_{5S}=1$ GeV$^{3/2}$, the latter figure approximating the  average value of
$I_{5s}(p)$ in the interval 0.5 GeV$\leq p\leq 1.2$ GeV. For comparison also
the actual values $J^{(1)}_{5S} (p_k)$  are given in the Table 1, obtained with
the $5S$ wave function, calculated in \cite{7} and approximated by $k_{max}=5$
oscillator functions. The widths $\Gamma_k$, calculated with $J^{(1)}_{5S}
(p_k) $ using
Eq.(\ref{1}) are given  in the last line of the Table 1.\\

{\bf Table 1.} Partial widths $\Gamma_k$ and overlap integrals
$J^{(1)}_{5S} (p_k)$ for six decay channels.
 \small{
\begin{center}
\begin{tabular}{|l|l|l|l|l|l|l|} \hline
$k$ &$B\bar B$& $ B\bar B^*$& $B^* \bar B^*$&  $B_s\bar B_s$&
$ B_s\bar B_s^*$& $ B_s^*\bar B^*_s$\\ \hline $p_k$,&&&&&&\\
GeV&1.26&1.16&1.05&0.84&0.68&0.48\\ \hline $M_k$, &&&&&&\\GeV &
5.28&5.30&5.32&5.37&5.39&5.41\\ \hline $Z_k$ &1&4&7&1/2&2&7/2\\
\hline $|\bar J_{5S}^{(0)}(p_k)| $&0.27&0.33&0.40&0.56&0.68&0.82\\\hline
$J_{5S}^{(1)}(p_k)$&-0.34&-0.44&-0.42&0.08&0.58&1.0\\\hline
$\Gamma_k,$&11&58&66&0.09&10&19\\ MeV&&&&&&\\
\hline

\end{tabular}

\end{center}}

\normalsize

As shown in \cite{5}, for $\Upsilon(4S)$ state the similar
computation yields $\Gamma_{tot} \simeq
\left(\frac{M_\omega}{2\omega}\right)^2 40$ MeV and comparison
with $\Gamma_{tot}^{exp} \approx 20$ MeV leads to the  estimate
$\left(\frac{M_\omega}{2\omega}\right)^2\approx  1/2$. As a result
one obtains
 $\Gamma_k$  for $k=1,...6$ as shown in bottom line of Table 1 with
 $\Gamma_{tot} \approx 150$ MeV.

 We note that the small width $\Gamma_4$ into $B_sB_s$ state is
 due to a nearby zero of $I_{5S} (p)$ and would be of the order of
 $\frac12 \Gamma_5 \approx 5$ MeV in the next approximation.

\section{ The $\pi BB$ decays of $\Upsilon (5S)$}

The $\pi B\bar B^*$ and $\pi B^*\bar B^*$ widths can be obtained
directly from the Eq. (71) of \cite{3} (see also Eq.(3) from
\cite{4}), $$ \Gamma_{5S} (\pi B\bar B^*) = \frac{M^2_{br}
(Z^*)^2}{N_c f^2_\pi} | J^{(1)}_{5 BB^*} (\vep, \vek)|^2\times
$$\be\times  \frac{d^3p}{(2\pi)^3} \frac{d^3 k}{(2\pi)^3} \frac{2\pi
\delta (E^{(0)}_5-E(\vep) -\omega_\pi(\vek))}{2\omega_\pi}.\label{6}\ee Here
$J^{(1)}_{5BB^*}(\vep, \vek)$ is the overlap matrix element with pion emission
defined in \cite{3,4,5,6}, namely \be J^{(1)}_{5BB^*} (\vep,\vek)=\int
\frac{d^3 q}{(2\pi)^3} \Psi_5 (c\vep-\frac{\vek}{2}+\veq) \psi_B (\veq)
\psi_{B^*} (\veq-\vek).\label{7}\ee

Note, that the decay occurs in the $S$- state, hence no additional
momentum $p_i$ in (\ref{7}), in contrast to (\ref{3}), and $\bar
y_{123}\approx 1$.

Expanding $\Psi_5, \psi_B$ in oscillator wave function series, one
can conveniently separate $\vep$ and $\vek$ dependence and write
\be J^{(1)}_{5BB^*} (\vep, \vek) =
e^{-\frac{p^2}{\Delta_5}-\frac{k^2}{4\beta^2_2}}
I_{5,11}(\vep).\label{8}\ee

Finally the width can be written as  $$\Gamma_{5S} (\pi B B^*) =
(Z^*)^2 \left(\frac{M_{br}}{ f_\pi}\right)^2\times $$\be\times
\int^{\omega_{\max}}_{m_\pi}\frac{e^{-\frac{2p^2}{\Delta_5}-\frac{k^2}{2\beta^2_2}}}{4\pi^3
N_c} | I_{5,11}(\vep)|^2\tilde M pk d\omega_\pi\label{9}\ee and
conservation law
$$\frac{p^2}{2\tilde M} + \omega_\pi =\Delta E= 0.255 ~{\rm
GeV};~~ \tilde M=2.65 ~{\rm GeV},$$ and $\omega_{max} = \Delta E,$
$\tilde M$ is the reduced mass of $B\bar B^*$.

The $\pi B^* \bar B^*$ decay width is obtained by replacing
$Z^*\to Z^{**}$ and $\Delta E = M (5S) - 2 M_{B^*}=0.21$ GeV. To a
good accuracy $Z^*\cong Z^{**}=1$, and one obtains  $$\Gamma_{5S}
(\pi BB^*)=\left( \frac{M_{br}}{f_\pi}\right)^2 15~{\rm keV},$$
\be \Gamma_{5S}(\pi B^* B^*) =\left( \frac{M_{br}}{f_\pi}\right)^2
3.3~{\rm keV}\label{10}\ee

Defining $\left( \frac{M_{br}}{f_\pi}\right)^2$ from dipion
widths, one has an estimate $\left(
\frac{M_{br}}{f_\pi}\right)^2\approx 1.5\div 2$, leading to the
predictions
$$\Gamma_{5S} (\pi B B^*)\approx (23\div 30) ~{\rm keV},$$
$$\Gamma_{5S} (\pi B^* B^*)\approx(5\div 6.6) ~{\rm keV}$$

\section{Results and  discussion}

We start with the results for the $BB$ decays, shown in Table 1.
Estimating $ \left(\frac{M_\omega}{2\omega}\right)^2$ from
$\Gamma_{tot} (4S)$ one finds that our calculated $\Gamma_{tot}
(5S)$ is approximately 150 MeV to be compared with the
experimental value from \cite{2}, $\Gamma^{\exp}_{tot} (5S) =
(110\pm 13)$ MeV. One can see a reasonable $\sim 25\%$ agreement,
taking into account a strong sensitivity to the actual form of the
$5S$  wave function.

An experimental hierarchy among the channels 1-3 is given by
relations \cite{2}.

\be \frac{\Gamma^{\exp}_1}{\Gamma^{\exp}_2}<0.92;~~
\frac{\Gamma^{\exp}_1}{\Gamma^{\exp}_3}<0.3;~~
\frac{\Gamma^{\exp}_2}{\Gamma^{\exp}_3}=0.324, \label{10a}\ee
while for $B_sB_s$ channels one has \cite{2} \be
\frac{\Gamma_4^{\exp}+\Gamma_5^{\exp}+\Gamma_6^{\exp}}{\Gamma_{tot}}=0.16\pm
0.02\pm 0.058\label{11}\ee and also \be
\frac{\Gamma^{\exp}_4}{\Gamma^{\exp}_6}<0.16;~~
\frac{\Gamma^{\exp}_5}{\Gamma^{\exp}_6}<0.16.
 \label{12}\ee
 Comparing wit our calculated $\Gamma_k$ in Table 1 one concludes,
 that all relations (\ref{10a})-(\ref{12}) are satisfied except for
 the last ones in  (\ref{10a}) and (\ref{12}).

 The situation is better for the averaged widths $\bar \Gamma_k$
 obtained from $\bar J^{(0)}_{5S}$. It is clear that more efforts
 are needed both on the theoretical side (improvement of wave
 functions) and on the experimental side (improvement of accuracy
 of $\Gamma^{\exp}_k$).

 As a whole, we have explained  the points 2) and 3) of
 Introduction, obtaining large $\Gamma_{tot}\approx
 \Gamma_{tot}^{\exp}$ and an approximate experimental hierarchy
 among the $\Gamma_k$, $k=1,...6$ In addition, we have calculated
 for the first time the $B\bar B^*\pi, B^*\bar B^*\pi$ widths of
 $\Upsilon (5S)$, which are as large as $O(10~ $keV).  This
 magnitude makes it possible to observe the $BB\pi$ decays by the
 Belle and BaBar Collaborations.

 On theoretical side  the  large factor of  $10^3$
 between $\Gamma_{5S} (BB)$ and $\Gamma_{5S} (BB\pi)$ is due to
 different vertex masses $M_\omega$ and $M_{br}$ respectively, and
 experimental confirmation of the magnitude of
 $\Gamma_{5S}(BB\pi)$ is important for understanding of quark pair
 creation mechanism. In a less direct way,  $M_\omega$ and
 $M_{br}$ yield the values of $\Gamma_{\pi\pi} (nS)$  and explain
 the large ratio $\frac{\Gamma_{\pi\pi}(5S)}{\Gamma_{\pi\pi}(4S)}
 \approx O(10^3)$, as it is discussed in \cite{5,6}.

The authors are grateful to M.V.Danilov, S.I.Eidelman for constant
support, suggestions and criticism, to P.N.Pakhlov and all members
of ITEP experimental group for stimulating discussions. The
financial support of  grants RFFI  06-02-17012,  06-02-17120 and
NSh-4961.2008.2 is gratefully acknowledged.

 \end{document}